# Title: QUANTUM ANALOG OF VIBRATION ISOLATION: From Room-Temperature Superfluorescence to High-Temperature Superconductivity

**Authors:** Kenan Gundogdu[1*], Franky So[2], Mark L. Brongersma[3], Melike Biliroglu[1], Gamze Findik[1]

**Affiliations:**

[1]Department of Physics, North Carolina State University, Raleigh, North Carolina 27695, United States

[2]Department of Materials Science and Engineering, North Carolina State University, Raleigh, North Carolina 27695, United States

[1,2]Organic and Carbon Electronics Laboratories (ORaCEL), North Carolina State University, Raleigh, North Carolina 27695, United States

[3] Geballe Laboratory for Advanced Materials, Stanford University, Stanford, CA, 94305, USA

*Correspondence to: kgundog@ncsu.edu

**The development and the use of quantum technologies are hindered by a fundamental challenge: Quantum materials exhibit macroscopic quantum properties at extremely low temperatures due to the loss of quantum coherence at elevated temperatures. Here, based on our recent discovery of room temperature superfluorescence in perovskites, we present the *Quantum Analog of Vibration Isolation, "QAVI",* model and explain how it protects the quantum phase against dephasing at high temperatures. We then postulate the requirements for observation of macroscopic quantum phenomena at practical temperatures and propose**

**a unified model for all macroscopic quantum phase transitions. We further present the general features of the temperature/density phase diagram of macroscopic quantum phase transitions that are mediated by the QAVI process and identify the similarities observed in the phase diagram of high-Tc superconductors. Understanding this fundamental quantum coherence protection mechanism is imperative to accelerate the discovery of high-temperature macroscopic quantum phenomena, and offers significant potential for developing quantum technologies functioning under practical conditions.**

In the 1940s, computing and telecommunication equipment were based on discrete electronic components such as vacuum tubes, resistors, capacitors, and inductors. A true transformation of these technologies into their current form required breakthrough discoveries in semiconductor devices and integration of these devices in tiny chips (See Fig. 1). Following the example of developments in electronics in the mid-20th century, quantum technologies promise a paradigm shift in every aspect of advanced technologies ranging from computation to communications. Similar to earlier discoveries in semiconductors, breakthrough developments in quantum materials are imperative to advance quantum technologies today.

Quantum devices rely on the ability to manipulate both wave and particle-like properties of physical objects. Because quantum properties are mostly observed in the smallest (atomic or molecular) form of matter, detecting these quantum particles and controlling their wave-like properties such as superposition, and entanglement using external probes are extremely challenging. Alternatively, quantum technologies can be built on macroscopic quantum phenomena such as superconductivity, Bose-Einstein condensation, and superfluorescence[1-6].

Materials exhibiting such macroscopic quantum properties feature more accessible physical dimensions that can be manipulated and measured by external controls and probes. For this reason, they can function as the fundamental building blocks in quantum technologies. However, for almost all cases, these macroscopic quantum phenomena are only observed at cryogenic temperatures[7-9] or under extreme pressure[10-12] limiting their practical applications. For instance, the state of the art in quantum computers based on superconducting qubits is only operational at sub-Kelvin temperatures[13]. Therefore, a grand challenge is to discover novel materials exhibiting macroscopic quantum phenomena at high temperatures and use these materials for quantum device components.

**Quantum Oscillator Phase Stability Challenge**

Since macroscopic quantum phenomena in their current state are not practical for quantum applications, we need to transform exotic macroscopic quantum phenomena into ordinary quantum phenomena that are observable under practical conditions. Here, we describe a generic macroscopic quantum system as an ensemble of coherent quantum oscillators. A macroscopic coherence is achieved when these oscillators sustain a common phase. Depending on the nature of the quantum oscillators, the resulting macroscopic quantum state exhibits exotic properties such as superconductivity, superradiance, superfluidity, and Bose-Einstein condensation[14-18].

Classical and quantum oscillators in the same medium can synchronize due to their collective interaction with their surrounding medium reaching a common phase. Classical examples include the synchronization of metronomes on the same platform, the orbital motion in the planet system, and the flashing of fireflies in the wild[19-23]. Quantum examples include the second-order quantum phase transitions described above. In both classical and quantum realms, the major obstacle to synchronization is the *phase stability* of oscillators[24].

Up until now, no theory or model was providing guidance to identify or design materials and/or systems exhibiting high-temperature macroscopic quantum phenomena. In our recent studies, we discovered lead-halide perovskites exhibit superfluorescence at surprisingly high temperatures[25,26]. Our investigation of this macroscopic phenomenon at unusually high temperatures led to the development of a **"Quantum Analog of Vibration Isolation (QAVI)"** model. In this perspective paper, we present the QAVI model in detail and explain its implications beyond superfluorescence. We specifically discuss the density/temperature phase diagram of QAVI mediated macroscopic phase transitions, which has a dome-like shape as commonly observed in high-$T_C$ superconductors[27,28]. The ideas presented here will help establish a set of design rules for novel quantum materials and devices and potentially lead to a transformation of quantum technologies similar to that of electronics technology in the 20$^{th}$ century.

**Quantum Analog of Vibration Isolation- "QAVI"**

To understand QAVI, it is insightful to start by understanding vibration isolation in a classical system. *In classical mechanics, to protect a system from ambient disturbances, instead of cooling the system to low temperatures, the system is isolated from the environment using a low-pass damped spring. In the QAVI model, similar protection is provided by a "low-pass quantum spring" such that the quantum oscillators are isolated from the ambient disturbances.*

To understand the dephasing mechanism of a quantum oscillator, we will use the particle in a box as an illustration. Any quantum oscillator, whether it is a Cooper-pair, an exciton, an electronic dipole, or a spin, can be considered as a particle in a box. For simplicity, Fig. 2 shows such a particle in a one-dimensional (1-D) box potential. The solution to Schrodinger's equation reveals the energies and wavefunctions of such a particle.

$$H\psi_n = E_n\psi_n \tag{1}$$

For a general quantum oscillator as the one in Fig. 2, the energy levels of the wavefunctions are determined by the potential energy "$V(x)$" term in the Hamiltonian. The solutions to Schrodinger's equation are standing waves with frequencies $\omega_n = \frac{E_n}{\hbar}$. Figure 2 shows two of these confined states and their superposition (pink wave packet) $\psi_1 + \psi_2$ making a quantum oscillator, which will periodically move between the boundaries of the box with a frequency of $\omega_{12} = \omega_2 - \omega_1$. The stability of the energy levels, $E_n$, determines the coherence lifetime. For electronic states in a condensed matter, these energy levels are determined by the potential and hence by the atomic positions. Therefore, any stochastic changes in the atomic positions induced due to the interaction with phonons result in dephasing, which becomes more severe at high temperatures. The conventional approach to deal with this problem in solid-state is to reduce the phonon population by cooling the solid and freezing the positions of the atoms.

**"Quantum Analog of Vibration Isolation - QAVI"** extends the quantum coherence without the need to lower the temperature. In QAVI, the electronic potential is isolated from the ambient phonons by some form of a "quantum isolator spring" (Fig. 2b). In solids, because thermal phonons are atomic motions, i.e. lattice modes, the "quantum isolator spring" has to be a lattice mode. In such a configuration, thermal phonons interact with the quantum oscillator through the lattice mode (quantum isolator spring) that it is bounded with. Similar to a classical low pass filter, this lattice mode will filter out the phonon disturbances depending on their frequency. If a thermal phonon mode has a higher frequency than the natural frequency of the quantum isolator spring (low-frequency lattice mode), its impact on the quantum oscillators is attenuated. The coupled electron-lattice mode combination makes a **polaronic quasi-particle**, and its formation is the key to the observation of high-temperature quantum phenomena[25,26].

Fig. 2b is a simplified schematic of a complex problem. A more complete representation of the QAVI model in extended quantum systems such as solids requires electronic band structure and phonon dispersion data. For extended materials, the calculations of polaronic effects in band structure are highly sophisticated and are not within the scope of this perspective article. Here, to provide an intuitive picture of QAVI in an extended solid, we consider the simplest system, a 1-D monoatomic chain. In this system, electronic bands form due to a periodic lattice potential, $V(x) = V(x + a)$, where $a$ is the lattice constant. Figure 2c shows the typical band structure represented in the 1$^{st}$ Brillouin zone, extending from $-\frac{\pi}{a} < k \leq \frac{\pi}{a}$, where $k$ is the crystal momentum and $a$ is the lattice constant. Figure 2d shows the phonon dispersion curves for the same system (Since we considered a monoatomic chain only acoustic modes are included. But these results can be extended over other structures). It is important to note that the phonon modes that are particularly responsible for dephasing are those close to the edge of the Brillouin zone, i.e, $k \approx \pm\frac{\pi}{a}$ in Fig. 2d. Because these modes lead to antisymmetric stretch, in other words, maximum compression and stretch in two adjacent unit cells. Hence, particularly those phonon modes cause large energy fluctuations and quick dephasing of electronic excitations. In contrast, the phonon modes that are around $k \approx 0$, are less destructive for the electronic phase. Because these modes extend over larger spatial distances compared to the lattice constant, their effect on the interatomic potential spacing is much smaller. In a structure without polarons, the phonon modes with $k \approx \pm\frac{\pi}{a}$ gets thermally populated as temperature rises and hence the electronic dephasing rate increases. In contrast when a large polaron forms, electrons are bound to a low energy lattice mode with a wavevector "$q = 2\pi/\lambda_p$", where $\lambda_p$ is the spatial extend of the polaron. This lattice mode of the polaron is sustained by the presence of electrons. In other words, it is a perturbation to the system with a periodic potential $V(x) = V(x + \lambda_p)$. Similar to the formation of bandgaps due to the periodic lattice potential, this new periodic perturbation to the potential will cause the formation of new pseudogaps within the Brillouin zone in positions at $\pm\frac{\pi}{\lambda_p}$, as shown by red lines in Figs 2c and 2d.

We call these new gaps "*pseudogaps*", because they are formed due to polaron formation and they only exist as long as the polaron is stable. In this configuration, the thermal population of phonon modes with higher frequencies than the polaron lattice mode (i.e, $\omega_p$, of the "$q = 2\pi/\lambda_p$" mode) is suppressed by these gap formations. This can be further understood if we analyze the oscillations classically as in Fig 2e. If modes around $k \approx \pm\frac{\pi}{a}$ starts to populate, they will try to move atoms with a higher frequency than the polaron lattice mode (i.e, $\omega_p$, of the "$q = 2\pi/\lambda_p$" mode). However, since the polaron has a lower natural frequency, it will respond to these high-frequency oscillations with a $\pi$ phase shift, and the phonon modes with frequencies beyond $\omega_p$ are suppressed, leading to an extended dephasing time.

In such a quantum system the polaron ***stability*** and the ***frequency of the lattice mode*** (and associated crystal momentum) are the two important characteristics to achieve a macroscopic quantum state at high temperatures. First, the polaron "***stability***" mostly depends on the binding energy of the lattice mode and the quantum oscillator making a polaron. If the binding energy is less than the thermal energy, *kT*, then the polaron is not stable and cannot provide the quantum isolation. In addition to the binding energy, the density of the polarons is another critical factor determining the stability. When the polaron density is too high, these quasiparticles become unstable due to the screening of the Coulomb force keeping the polaron intact[29,30]. Hence, QAVI can only protect the quantum phase of an oscillator at specific temperatures and densities such that the polaron is stable. Second, the "***frequency of the lattice mode***" depends on the atoms in the lattice mode. Since QAVI acts as a low-pass filter, polarons with heavier atoms provide stronger isolation. In other words, a polaron formed by a quantum oscillator strongly bound to a low energy lattice mode is required to provide a good quantum phase protection at high temperatures.

**Macroscopic Quantum Phenomena**

Macroscopic quantum coherence arises due to the synchronization of an ensemble of oscillators placed in the same medium. It is important to note that synchronization is a spontaneous process that is commonly observed in ensembles of particles at many different scales. A good classical example is metronomes on a platform. The collective interaction of randomly oscillating metronomes with their medium and each other leads to a coherently oscillating metronome ensemble[21]. Similar to metronomes on the same platform, quantum particles also tend to synchronize under certain conditions. When this happens, the system forms a macroscopic quantum state with exotic properties such as superconductivity, Bose-Einstein condensation, superfluidity, and superfluorescence. For instance, in the case of superfluorescence, the synchronization of the dipoles is driven by vacuum field fluctuations and leads to collective radiation of the ensemble by emitting a burst of photons[31]. In a quantum system, we generally call this process a symmetry-breaking-phase transition. But in this perspective article, we will use the terminology of "synchronization" as it refers to the actual event happening over time.

*The necessary condition for macroscopic coherence is to have a synchronization time shorter than the phenomenological dephasing time*. The *"synchronization time"* depends on the density of oscillators. When the oscillator density is increased, the synchronization time is reduced[7,14,32]. In a classical system, simulations based on the Kuramoto model clearly show this tendency[33]. For quantum systems, superfluorescence is a good example, where the time for synchronization decreases with $lnN/N$, where $N$ is the density of oscillators[7,8,34].

Figure 3 shows an illustration of the competition between synchronization and dephasing in a timeline. Assuming that the dephasing time is fixed at a certain temperature, the synchronization time is determined by the density of oscillators. If the density is below a critical value, then synchronization of the system will not take place, and the ensemble remains incoherent. As the density is increased close to a threshold value, synchronization will start competing with dephasing leading to instability and strong fluctuations in the system. When the density of the oscillators is significantly higher than the threshold value, the system is completely synchronized forming a macroscopically coherent state. In a typical condensed matter system, the electronic dephasing times are a few femtoseconds at practical temperatures. Hence the required density of oscillators for faster synchronization than dephasing is physically impossible to achieve at high temperatures. Therefore, it is important to note that we need to extend the dephasing time such that quantum oscillators can synchronize before dephasing. Our superfluorescence studies on perovskites show that QAVI extends the dephasing time by several orders of magnitude enabling the formation of a macroscopic coherent state even at room temperature[26].

## "Universality of Macroscopic Quantum Phase Transitions"

With our understanding of the QAVI protection mechanism, we postulate that this model might be used as a universal picture to understand other macroscopic quantum phenomena and provides conditions for observing them at high temperatures. To observe a macroscopic phase transition, the quantum oscillators should be bound to a lattice-mode forming polaron. The lattice mode should preferably involve the lowest possible energy mode that can be coupled to the electronic oscillators. Finally, there is a critical range of quantum oscillator density at which synchronization can be achieved; at too small densities the synchronization process is too slow, and at too large densities the polaron is not stable.

In this universal picture, the oscillator density and the phenomenological dephasing rate determine the temperature/density phase diagram of the formation of a macroscopic quantum state. In Fig. 4, we illustrate such a phase diagram. For a given temperature, the quantum coherent state survives between two threshold density points. The lower bound density is determined by the dephasing rate. Here, the ensemble needs to have enough density of oscillators for a faster synchronization than dephasing. The upper bound density is determined by the polaron stability. Beyond the upper bound density, the polaron is unstable, and hence there is no QAVI protection. As the temperature increases, the range of polaron densities that can sustain the macroscopic coherent state is reduced. This is both due to the increase in the threshold density for synchronization and also the decrease in the upper bound density due to polaron instability. There is a critical maximum temperature at which the density of oscillators required for synchronization is the same as the density of oscillators in that the polaron is unstable. Beyond this critical temperature, macroscopic coherence cannot form. Therefore, the temperature/density phase diagram for all macroscopic quantum phenomena mediated by the QAVI process will have a dome-like shape.

Interestingly, such a dome-shaped phase diagram is commonly observed in high-$T_c$ superconductors [35,36]. This phase diagram indicates that for a given temperature, superconductivity starts at a threshold doping density (known as a quantum critical point) and vanishes at a high doping density. In superconductors, one of the proposed mechanisms for pairing involves bipolarons with two opposite spin electrons bound to a lattice mode[37], which is similar to electron-hole (dipole) binding to a lattice mode in our observation of high-temperature superfluorescence[25,26]. We propose that the loss of superfluorescence at high excitation densities, and the loss of superconductivity at high doping densities might share a common origin. At high doping densities, the instability of bipolaron detrimentally affects both QAVI protection and the pairing mechanism. Hence the so-called quantum criticality that is under the superconducting dome might be due to the synchronization of quantum oscillators, and the loss of superconductivity may be due to the loss of QAVI protection at high carrier densities. In superconductivity, measurement of the synchronization process is experimentally challenging, whereas in superfluorescence the synchronization dynamics are easily measurable using time-resolved spectroscopy. In our studies, we clearly showed that once the polaron becomes unstable, the superfluorescence vanishes[26]. To verify the universality of the dome-shaped phase diagram, we experimentally measured the superfluorescence in phenethylammonium cesium lead bromide/chloride (PEA:CsPbBr$_{2.5}$Cl$_1$) quasi-two-dimensional (2-D) hybrid perovskites at various temperatures and excitation fluences. Strikingly, the results presented in Fig. 4b clearly show such a dome-shaped phase diagram for superfluorescence. *Hence, we provide a piece of significant evidence that the QAVI mechanism may underlie the formation of macroscopic quantum coherence in both superconductivity and superfluorescence.*

Another important property of the materials that exhibit QAVI is the pseudogap formation. As we described above, when an electron binds to a particular lattice mode with a wavevector $q = 2\pi/\lambda_p$, the periodic modulation of the electronic potential due to the particular lattice mode results in new electronic band gaps at $\pm\frac{\pi}{\lambda_p}$. Since polaron formation depends on temperature and excitation density, the pseudogap formation can be also plotted on the same phase diagram. At the lowest excitation densities ($N \approx 0$), polarons are stable at temperatures, $T \leq E_b/k$, where $E_b$ is the polaron binding energy and $k$ is the Boltzmann constant. As the density is increased, Coulomb screening further destabilizes the polaron. Therefore, polarons and pseudogaps are observable on the left side of the dashed line and inside the dome in Figure 4a.

Last but not least another important characteristic of the high-T$_c$ superconductor phase diagram is the charge fluctuations that take place at the transition doping density at different temperatures[38-40]. A similar fluctuation in superfluorescence intensity has been observed by us and others, indicating that these fluctuations are due to the competition between synchronization and dephasing[26,41-43]. Since at the threshold excitation fluence a stable coherence cannot be sustained, the system fluctuates between the macroscopic superradiant state and the incoherent state. These similarities between high-Tc superconductors and high-temperature superfluorescence indicate that there are some one-to-one correspondences between the two different macroscopic quantum phenomena, suggesting both processes might share the same QAVI protection mechanism.

In conclusion, we used the QAVI model to explain how macroscopic quantum states can form at high temperatures. In this model, the phase of a quantum oscillator is protected by strongly bound lattice modes, suppressing dephasing due to thermal noise with a process very similar to vibration isolation in classical systems. The proposed QAVI protection mechanism might provide insight into the understanding of other high-temperature quantum phenomena including superconductivity. More importantly, QAVI is a general process that has implications beyond known macroscopic quantum phase transitions. Here we discussed electronic excitations confined in a lattice with Coulombic potential. But similar protection mechanisms can be developed for other quantum oscillators including but not limited to those that involve photons in cavities, and spins in magnetic potentials and facilitate their coherent behavior. This new understanding provides an intuitive picture and an opportunity for designing new materials tailored for various quantum effects at high temperatures. QAVI model will open up a new direction for quantum materials and quantum device research as it will accelerate the discovery of macroscopic quantum phases in tailored structured materials.

**Acknowledgments:**
This work is supported by the National Science Foundation Designing Materials to Revolutionize and Engineer our Future program (grant #1729383) and the NC State University Research and Innovation Seed Funding (RISF). **Author contributions:** K.G. conceived the QAVI idea and its implication for macroscopic quantum phase transitions and phase diagrams presented here and drafted the manuscript. K.G and M.L.B then developed the band picture related to QAVI. M.B and G.F prepared the illustrations in figures and performed the experiments in Figure 4b. F.S provided the samples and edited the manuscript. All authors contributed to writing the paper. **Competing interests:** The authors declare no competing financial interests. Correspondence and requests for materials should be addressed to K.G.

**Figures**

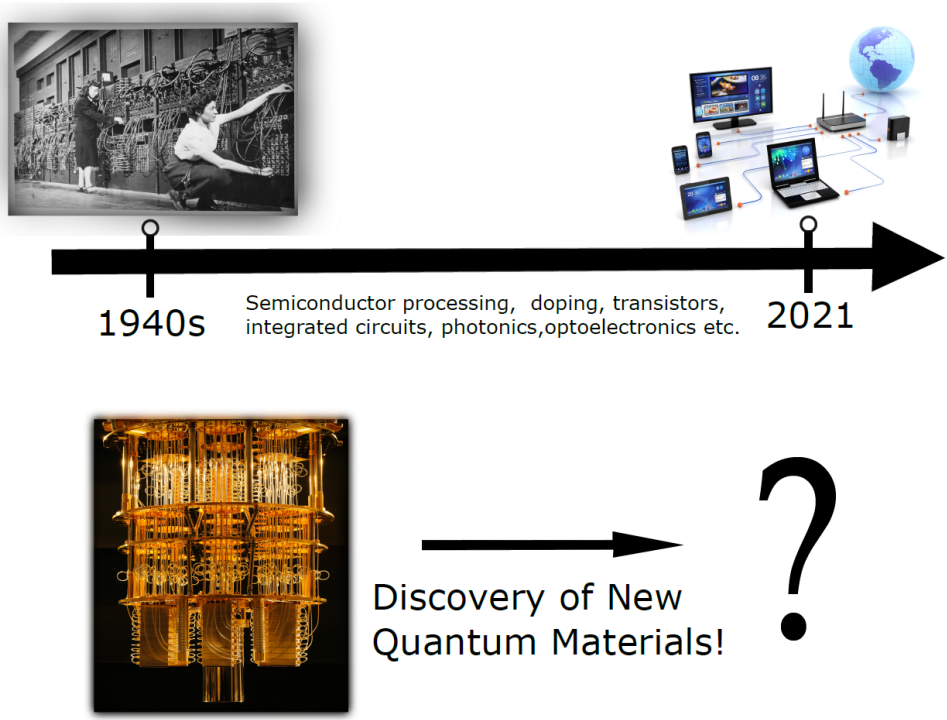

**Fig.1: Historical progress of electronics technology.** The panel above shows the first ENIAC computer built-in 1945. The discoveries took place over the last century in the electronic materials and devices such as processing of semiconductors, transistors, integrated circuits transformed electronics technology to the contemporary electronics that became integrated part of our life. The bottom panel shows the quantum computer functions at sub-Kelvin temperatures. The observation of macroscopic quantum phenomena at cryogenic temperatures limits widespread utilization of quantum technologies in consumer markets. Discovery of materials that exhibit macroscopic quantum states at high temperatures and invention of room temperature devices that manipulate macroscopic quantum states will lead to transformation of quantum technologies that is similar to what we have observed in conventional electronics in the last 80 years. Images: Photograph of World's First Computer, the Electronic Numerical Integrator and Calculator, National Archives **(**306-PSE-79-120); Home solutions and wifi devices network**,** pictafolio/Getty Image; IBM quantum computer by IBM Research/Graham Carlow Photography (CC BY-ND 2.0).

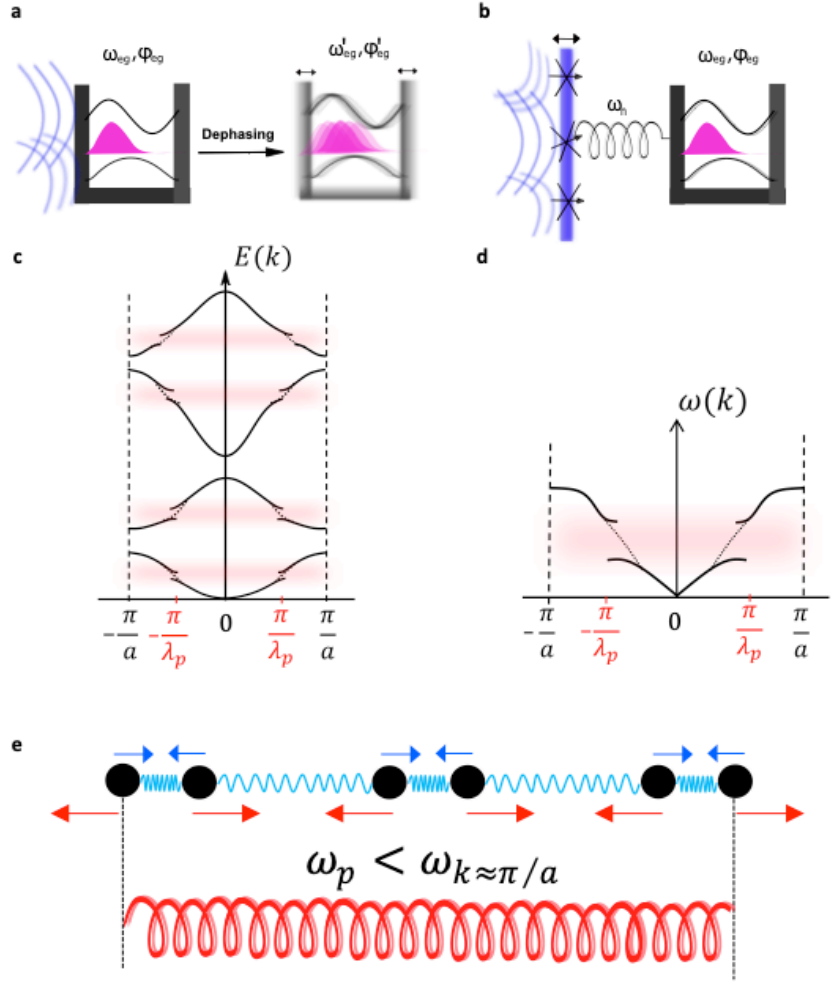

**Fig. 2: Illustration of QAVI. a,** A quantum oscillator (magenta wave packet) is formed by superposition of two confined states in a box potential. This state oscillates left and right in the box with a frequency of $\omega_{12}$ and acquires phase "$\varphi_{12}$" over time. The stochastic changes in the atomic positions due to the interaction with phonons (blue waves) changes the oscillation frequency randomly and causes dephasing. **b,** The box potential is protected from the environmental disturbances by some form of a "quantum spring" similar to classical systems. The noise will be filtered out depending on the relative frequency compared to the frequency of the spring ($\omega_s$). **c,** The electronic band structure for 1-D monovalent monoatomic chain. The bandgaps and band widths are primarily determined by the periodic lattice potential. **d,** The corresponding phonon dispersion curves. In **c and d**, the red regions are the pseudogaps formed due to electron-lattice mode coupling in both electron bands and phonon bands. For simplicity we only showed one pseudogap per branch. Depending on the periodicity of the polaron potential multiple gaps can open within a branch. **e,** The illustration of vibrational isolation in a 1D chain segment. Blue waves represent a phonon mode that is close to the Brillouin zone edge. The red spring illustrates the lattice mode of the polaron. When blue modes are thermally excited, it will try to drive the polaron lattice mode. But because polaron lattice mode has a lower frequency, it will oscillate with a $\pi$ phase shift and suppress the high frequency thermal phonon modes.

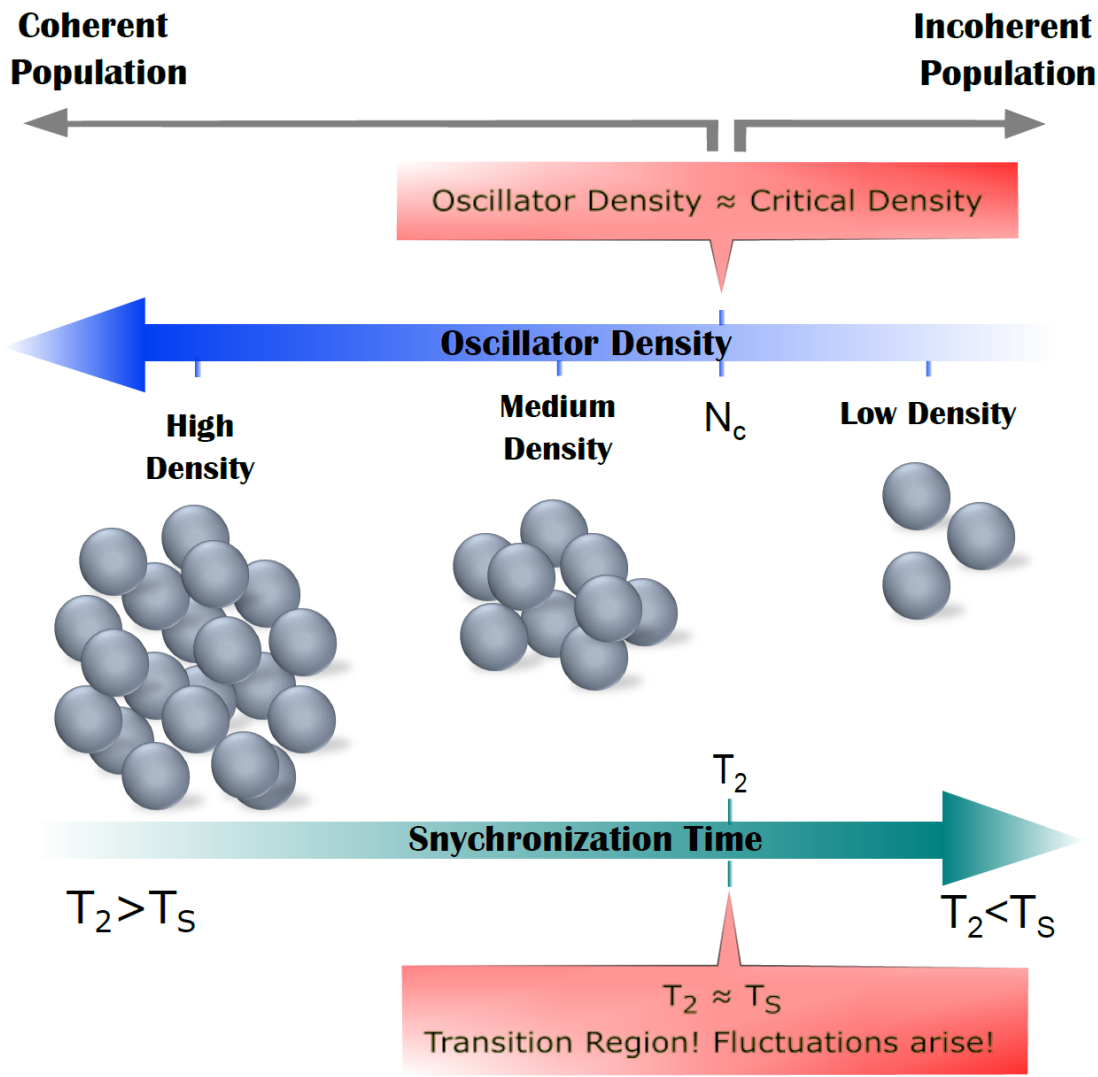

**Fig. 3: Graphical representation of macroscopic quantum state formation at a critical oscillator density.** At a given temperature, oscillators have a characteristic dephasing time $T_2$. If an ensemble of oscillators can synchronize before the dephasing time then a macroscopically coherent state forms, if not the system remains incoherent. The green arrow shows the timeline for synchronization "$T_s$". The blue arrow shows the density of oscillators. As the density increase, the synchronization time shortens. Below the critical oscillator density ($N_c$), the synchronization time $T_s$ is longer than dephasing time $T_2$. Therefore, macroscopic coherence cannot build up. Beyond the critical density, the system forms macroscopic quantum coherence. At the critical density, the competition between the dephasing and synchronization leads to fluctuations.

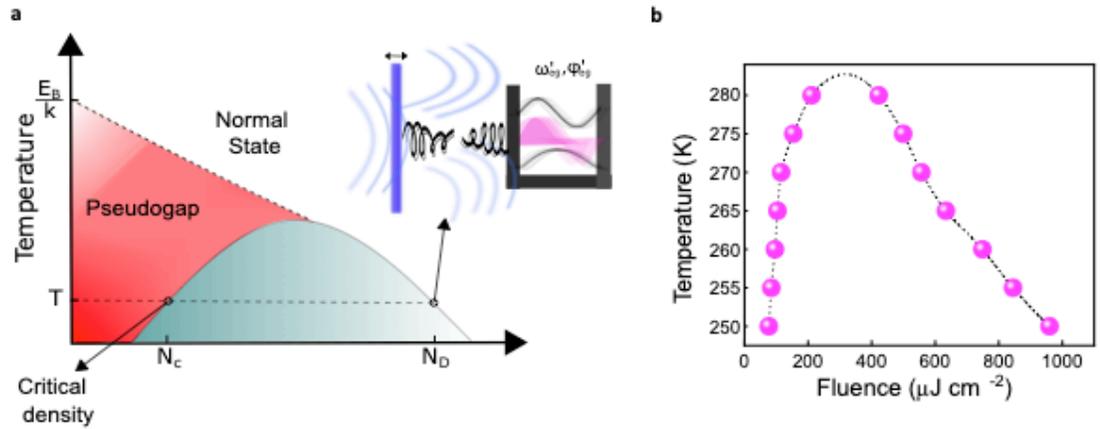

**Fig.4: A schematic representation of commonly observed dome shaped oscillator density-temperature phase diagram for high-$T_c$ superconductors. a,** The model presented in this paper predicts a similar dome shaped phase diagram for all macroscopic quantum phenomena and a pseudogap phase with the QAVI protection mechanism. The horizontal dashed line marks the phase transition conditions for a fixed temperature. The density of oscillators needs to be at a certain critical density ($N_c$), so that the synchronization can take place before dephasing. Since the dephasing time reduces with increasing temperature, the oscillator density must increase to realize the phase transition at higher temperatures. For a given temperature the macroscopic quantum state is sustained for a range of oscillator densities beyond $N_c$. But as the density is increased significantly, Coulomb screening can lead to destabilization of the polaron leading to loss of QAVI mechanism as represented with the quantum well with the broken spring. The density "$N_d$" at which the polarons are destabilized sets the upper limit of the density in the phase diagram. The red region is the pseudogap phase in which the electronic band structure exhibit pseudogaps due to polarons. In this phase he macroscopic quantum state do not form due to lack of synchronization. **b,** The temperature-fluence plot of the quasi-2D PEA: $CsPbBr_xCl_y$ thin films exhibits a dome-like phase diagram.